\documentclass[conference]{IEEEtran}
\IEEEoverridecommandlockouts
\usepackage{cite}
\usepackage{amsmath,amssymb,amsfonts}
\usepackage{algorithmic}
\usepackage{graphicx}
\usepackage{textcomp}
\usepackage{xcolor}
\def\BibTeX{{\rm B\kern-.05em{\sc i\kern-.025em b}\kern-.08em
    T\kern-.1667em\lower.7ex\hbox{E}\kern-.125emX}}
\begin{document}

\title{On-board Electrical, Electronics and Pose\\ Estimation System for Hyperloop Pod Design 
\\}

\author{\IEEEauthorblockN{Nihal Singh*, Jay Karhade*, Ishika Bhattacharya*, Prathamesh Saraf*, Plava Kattamuri*, Alivelu Manga Parimi}
\IEEEauthorblockA{
Birla Institute of Technology and Science, Pilani\\
Email: nihal.s.singh@gmail.com, jaykarhade@ieee.org, ishika2602@gmail.com,\\
pratha1999@gmail.com, plavakattamuri@gmail.com, alivelu@hyderabad.bits-pilani.ac.in}
*All authors have equal contribution}
\maketitle

\begin{abstract}
Hyperloop is a high-speed ground-based transportation system utilizing sealed tubes, with the aim of ultimately transporting passengers between metropolitan cities in efficiently designed autonomous capsules. In recent years, the design and development of sub-scale prototypes for these Hyperloop pods has set the foundation for realizing more practical and scalable pod architectures. This paper proposes a practical, power and space optimized on-board electronics architecture, coupled with an end-to-end computationally efficient pose estimation algorithm. Considering the high energy density and discharge rate of on-board batteries, this work additionally presents a robust system for fault detection, protection and management of batteries, along with the design of the surrounding electrical system. Performance evaluation and verification of proposed algorithms and circuits has been carried out by software simulations using both Python and Simulink.
\end{abstract}


\begin{figure*}[htb]
    \centering
    \centerline{\includegraphics[width = 488pt]{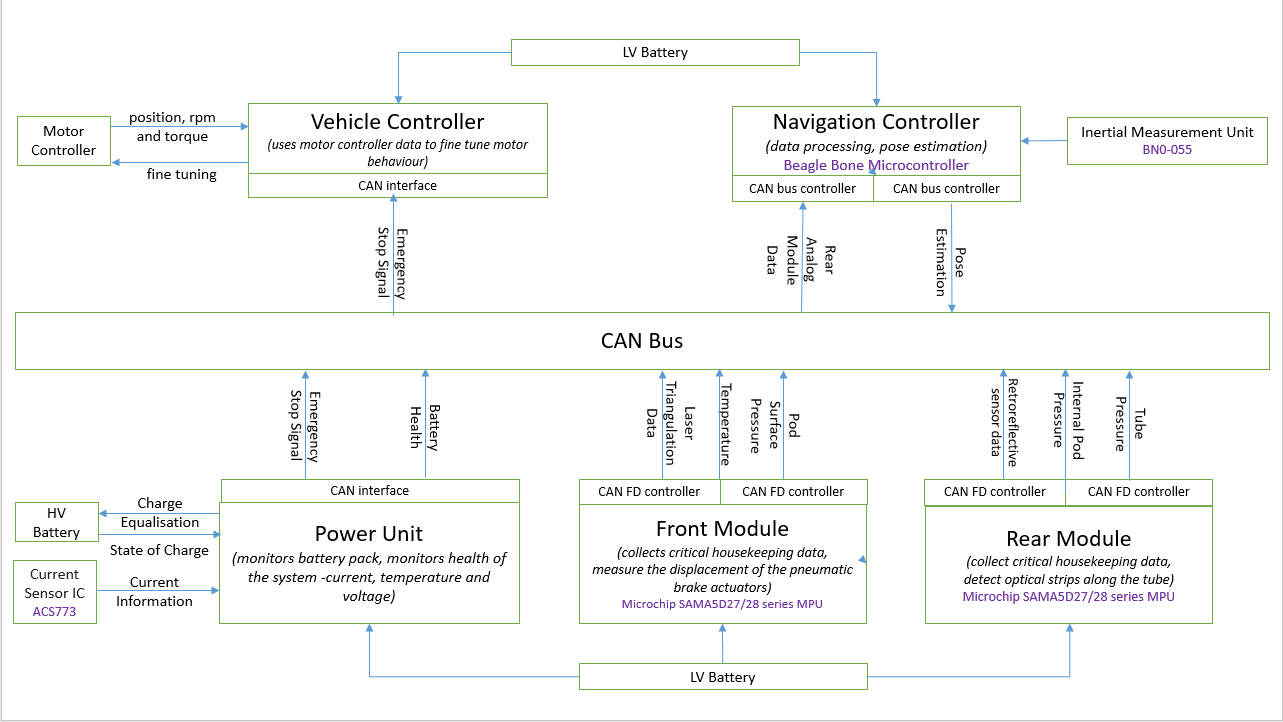}}
    \caption{On-board System Architecture Model}
    \label{fig:sysarch}
\end{figure*}

\section{Introduction}
The Hyperloop is a novel mode of high speed passenger and freight transportation, based on an open-source vactrain design released by a joint team from Tesla and SpaceX. It is a form of low friction ground transport aiming for speeds over 700mph with the help of partially evacuated low-pressure tubes. Work on air-pressure driven transportation dates back to the early 1800s with the introducton of the concept of an “atmospheric railway” ~\cite{medhurst1810} and coinage of the term Vactrains ~\cite{goddard}. Interest in the concept was renewed with Elon Musk’s original white paper on Hyperloop Alpha ~\cite{musk2013}. The emphasis on the need for more efficient transport systems is increasing in recent times, and there has been considerable evolution in the various technical aspects of the pod. 

The nervous system of any general pod design is comprised of its software and embedded systems. The expectations and challenges of this combined subsystem have been well-detailed in the paper on the Goose-3 pod~\cite{waterloo}. Using a divide-and-conquer approach, the tasks were divided between $(1)$ the embedded system, responsible for collecting data from on-board sensors and executing control panel commands, and $(2)$ the communication system, responsible for transmission of data between the remote control panel and the embedded system. Key features include the presence of a CAN-BUS and a Master-Slave Design with master and hub units. In the MIT design~\cite{mit}, the system is split into seven important blocks. The Master Computer serves as the connection between the remote laptop and embedded micro-controllers. A dedicated Flight Controller whose sole responsibility performs state estimation, and Front and Rear Analog modules read non-critical sensor data for monitoring. In addition, there is a fiducial detector module and a module for braking. The rLoop design~\cite{rloop}, too, uses a combination of retro-reflective photo-detection sensors and inertial measurement units for navigation. 

The heart of the system, encapsulated by the power node, focuses on power conditioning, battery monitoring, and overall system safety. Prolonged operation of the pod at maximum capacity, can, however, result in increased temperatures in the cells, which eventually may lead to a phenomenon known as electrical arcing.



To prevent arcing, the rLoop team~\cite{rloop} designed a vessel of carbon dioxide to store the battery pack,  whereas the Georgia Tech pod~\cite{georgiatech} built a water cooling based system for batteries and other electronics. Other teams like the one from UC Santa Barbara~\cite{ucsb} programmed their system for partial or complete shut down and execution of emergency protocols in the case of detected malfunctions.

Another important part of the hyperloop pod is the pose estimation in real-time. Generally, the parameters that are measured during pose estimation are acceleration, velocity and position and attitude. Pose estimation tracking is done for a variety of reasons ranging from safety to motion-planning and even vehicle control. A range of sensors are used for obtaining the pod’s pose, however, each of them suffer from their own drawbacks which has been discussed in the pose-estimation section further. This is facilitated by a robust set of position estimation algorithms. Effective pose estimation is constituted by a low latency system with high resolution and accuracy of the estimates. Nikolaev~\cite{waterloo}, in his work on the Waterloo pod’s software system, proposes a minimal Kalman filter that simply fuses the data from 3 Inertial Measurement Units (IMUs). Drift and possible existing correlation between the sensors due to sensing the same physical quantity of a moving body are some of the limitations of this method. The MIT Hyperloop Report~\cite{mit}, has proposed an Extended Kalman Filter(EKF) that takes into account ground truths by reading the fiducial markers throughout the tube length which not only serves as a comparison stage to the IMUs, but can act as a recalibration step at constant intervals thus reducing the bias and drift from true estimates even further. While this serves as a basic Kalman filtering update, it is required to analyze the velocity, acceleration and position models and introduce an element of independent sensing for each parameter. Alternate systems taken by other teams like UCI~\cite{uci} include the ellipse-N commercial model which is an out-of-the-box model for pose estimation. The ellipse-N series uses Global Positioning System(GPS)/Global Navigation Satellite System(GNSS) coupled with IMU sensing which while offers extremely accurate measurements and integrate some of the most accurate positioning algorithms may suffer inside the tube due to satellite position attenuation and failure~\cite{caronGPSIMU}. A recent patent~\cite{martinpatent} assumes passive elements throughout the tube length which would yield accurate and precise estimates, but it is not immediately clear as to what the economic costs would entail. The focus hence should be to develop a robust positioning technique that can utilize onboard and internal tube structure resources with minimal assumptions and external installations to provide accurate estimates.

This paper introduces a custom on-board hardware architecture for end-to-end interfacing and internal control of electrical and electronics components. The work presented in this paper further proposes 1) detailed position, velocity and acceleration estimation algorithms specific to Hyperloop pod systems, along with and an efficient electrical system. Additionally on Electronics System Design, the board-level architecture proposed leverages salient features of the aforementioned embedded system designs while prioritizing distributed sensing and control along with the efficient use of hardware. Often, in the interest of better performance, due consideration is not given to monitoring the health of the on-board power systems. 2) In the Electrical System Design, safety has been strongly emphasized on, with focus on arcing mitigation and fault-detection. This falls under the purview of the software subsystem, as does pose estimation. 3) For pose estimation, the proposed design focuses on developing a robust positioning technique that can utilize on-board and internal tube structure resources with minimal assumptions and external installations to provide accurate estimates. It makes use of a simple, but diverse, range of sensors namely: IMUs, Tachometers, Absolute Optical Encoders and Fiducial markers and simulated acceleration data following a pre-planned trajectory. It has been assumed for simplification of the algorithm, that the motion is along a linear axis. A Kalman filter is used for fusing estimates and the updation follows a hierarchical fashion updating the estimates in the order: Acceleration, Velocity followed by position. Attitude estimation is done in an independent loop and involves only IMUs. Each of these subsystems has been further elaborated on in the following sections. 

\section{Electronics System Design}
Sensing and monitoring of the pod's environment and internal functionalities is critical for a hyperloop pod. The data gathered by various sensors is processed to determine the core parameters for all on-board control and actuation. The presented system architecture model builds on the work of other cited literature to propose a distributed noise-immune design with minimal hardware overhead. Each subsequent section details the various components used in engineering this system as shown in Fig~\ref{fig:sysarch}, along with descriptions of their hardware layout and functionalities.

\subsection{CAN Bus}
The electronics system design is based on a Controller Area Network (CAN)-BUS shown in Fig~\ref{fig:sysarch}, popularly used in automobiles to simplify connections between different electronic control units. This reduces the chances of single point failures by making the system more noise-robust. Real-time, reliability and flexibility make it an indispensable network communication technology applied in the automobile network communication field ~\cite{CANBUSev}. It makes communication with all the units possible without direct signal lines. A single point of entry for the vehicle controller allows for central diagnostics. The CAN bus' use of differential signals ensures there is no noise in communication between critical systems and the vehicle controller. Messages on a CAN-BUS are given priorities. The highest priority signals, like the emergency stop signal, gain control of the bus in the shortest period of time.
\subsection{Primary Board (Navigation Controller)}

The navigation controller shown in Fig~\ref{fig:sysarch}, bears the brunt of computation - the position estimation calculations and data processing. Usage of a BeagleBone Black has been deemed optimal since it has two internal CAN-Bus controllers (DCAN0 and DCAN1). One BNO-055 (IMU) will be directly interfaced to this microcontroller, while the other one is interfaced to the rear analog module. 

\subsection{Front and Rear Modules}
\begin{figure}[htbp]
\centerline{\includegraphics[width = 244pt]{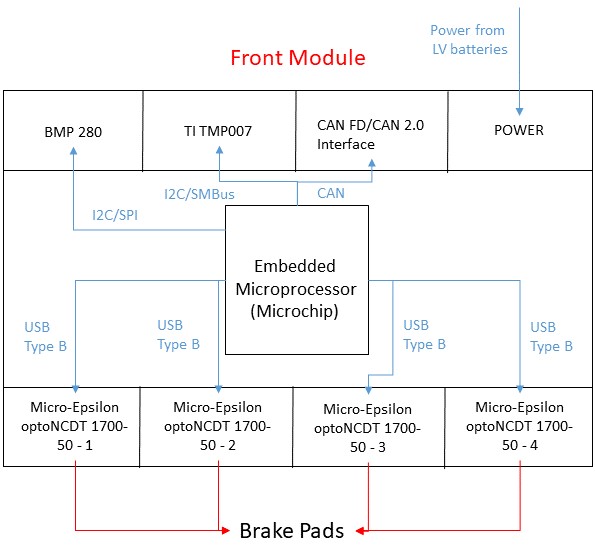}}
\caption{Board Layout proposed for design of Front Sensor Module}
\label{fig:frontmodule}
\end{figure}

\begin{figure}[htbp]
\centerline{\includegraphics[width = 244pt]{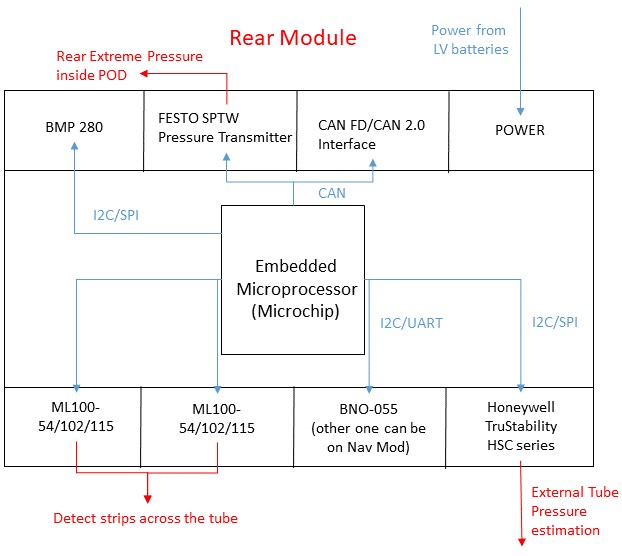}}
\caption{Board Layout proposed for design of Rear Sensor Module}
\label{fig:rearmodule}
\end{figure}

Two sensor modules as shown in Fig~\ref{fig:frontmodule} and Fig~\ref{fig:rearmodule} are used to interface directly to the sensors and act as a relay to transfer sensor captured information to the main controller. Based on the required positions of the sensors, it was decided to have the Front analog module placed near the brake pads and batteries, and the rear analog module placed behind the wheels.

The front analog module interfaces with four laser triangulation sensors (Micro-Epsilon optoNCDT 1700-50) used to accurately measure the displacement of the pneumatic brake actuators and provide important feedback data to facilitate further brake actuation. Along with this, the module also interfaces with an infrared thermopile sensor (TI TMP007) and a barometric pressure sensor (BMP 280). These two sensors measure the temperature and pressure inside the pod and relay on crucial housekeeping data which may trigger emergency modes and detect failures.

The Rear Analog module interfaces with both the Retro-reflective sensors (ML100-54/102/115) in order to detect the optical strips lined across the vacuum tube, which in turns helps in removing bias from the position estimation algorithm. A piezoresistive silicon pressure sensor (Honeywell TruStability HSC series) is also connected, and is used to determine the external tube pressure. For the rear end internal pressure estimation, a combination of BMP 280 and FESTO SPTW Pressure Transmitter has been used.

Both modules are fabricated around a Microchip SAMA5D27/28 series MPU. These are high performance ultra-low power Arm Cortex A5 CPU based embedded microprocessors with memory support. Additionally, each MPU supports two master CAN-FD (MCAN) controllers, and supports event triggered transmission. Both boards are supplied power from the LV batteries.

\subsection{Vehicle Controller}
The vehicle controller acts as the main flight controller of the pod.   It executes an Emergency Braking Protocol in case of any faults in the High-Voltage supply, damage to critical systems or any such unfortunate circumstances compromising the safety of passengers on board.
The vehicle controller requires only single CAN capability to interface with the motor controller. This interface is used to constantly poll information from the motor controller regarding the position, revolutions per minute (RPM) and torque control of the motors. This data will be used in a localized control loop to fine tune and accurately alter the behaviour of motors. 

\subsection{Power Node}
The main functionality of the Power Node is to keep a check on the health of the battery packs through a Battery Management System (BMS) and sensors checking the current, temperature and voltage. The BMS monitors features like the State of Charge of each cell, and charges/ discharges each cell accordingly. It has charge and discharge drivers for charge equalization. This is essential to provide cells from overcharging, and further damage as well. The BMS communicates with the microcontroller using I2C communication. The LM75A is a temperature sensor with an included Analog to Digital Converter and communicates with the power node with an I2C interface. The LM75A can provide digital temperature readings for a range of -55°C to 125°C.

The ACS773 is a Hall-Effect Based sensor which can detect currents sampled in a range of +/-100A. This is interfaced with an Analog to Digital Converter and then sent to the Power Node. It monitors current outside the Battery and to the other components, like the Motor Controller. 

\section{Electrical System Design}

The electronics system detailed in the previous section discusses the hardware needed for computation, control, and actuation, while the electrical system focuses on the design of an architecture used to facilitate power conditioning, distribution, and regulation. This section particularly focuses on discussing the design of the on-board DC to AC inverter model.

\subsection{DC to AC Inverter Model}
AC-powered linear induction motors are becoming a popular option for Hyperloop propulsion systems. The “Claude Nicollier” prototype from Swissloop ~\cite{swissloop} uses LIM based propulsion to reach a top speed of 252kmph, highlighting their potential. Linear induction motors work on a principle similar to a rotating squirrel-cage induction motor ~\cite{limsquirrelcage}. Hyperloop pods use a Double Sided LIM design ~\cite{dslim}, with two primaries face-to-face that interact with the rails, which form the long secondary. As part of future work on the design of our pod, LIMs are a serious consideration and in the same direction, a custom DC to AC inverter model has been developed to power the motors. The Simulink model for the same can be found in Fig~\ref{fig:simulink}.

\section{Battery Management System}
In order to satisfy power requirements, the proposed Hyperloop pod will house two different battery systems- a Low Voltage system for the sensors, boards, and other low voltage components and a High Voltage system to supply power to the motors. To power the pod, lithium iron phosphate batteries have been chosen as they are the safest of all lithium battery chemistries. A battery management system as shown in Fig. 4 is required to monitor these two in-house battery packs, as any faults in this system can be fatal to the pod and can cause arcing, compromising pod safety.   

\subsection{Arcing}
The high voltage batteries may experience temporary overloading, resulting in the ionization of surrounding air, in a phenomenon known as electrical arcing. This may happen when Hyperloop pods operate their motors at maximum capacity. Arcing causes a breakdown of the battery chemistry causing excessive electrical discharge through uncontrolled current conduction. The likeliness of an arcing failure may be further increased by gaps or breaks in battery insulation, corrosion, wear and tear on the batteries' exterior, and overloaded plug terminals. These arcs can cause overheating and even thermal runaway in the battery, leading to dangerous fires. 

\subsection{BMS Design for Hyperloop Pod}
To ensure arcing does not occur, it is important to use the appropriate Battery Management System (BMS) and have the necessary protection circuitry in place. The BMS keeps track of factors like total voltage, total current, voltage across each cell, temperature, pressure to avoid overcharge, overdischarge, or any other possible faults ~\cite{languangkeyissues}. Other main characteristics of this BMS include the prediction of the state of charge of the battery, ~\cite{SOCXiong}, state of health of the cell, and fault diagnosis for batteries, sensors, loose connections, and combustible gas levels. The BMS is designed to handle any faults by controlling the charge of cells, or switching off the battery power supply if necessary to obstruct damages and arcing as well. A cooling system using inert gases, carbon dioxide, or heptafluropropane can be designed to prevent arcing. These obstruct any flammability or re-ignition of arcs that may be generated in the battery pack~\cite{Cooling}.
\begin{figure}[htbp]
\centerline{\includegraphics[width = 244pt]{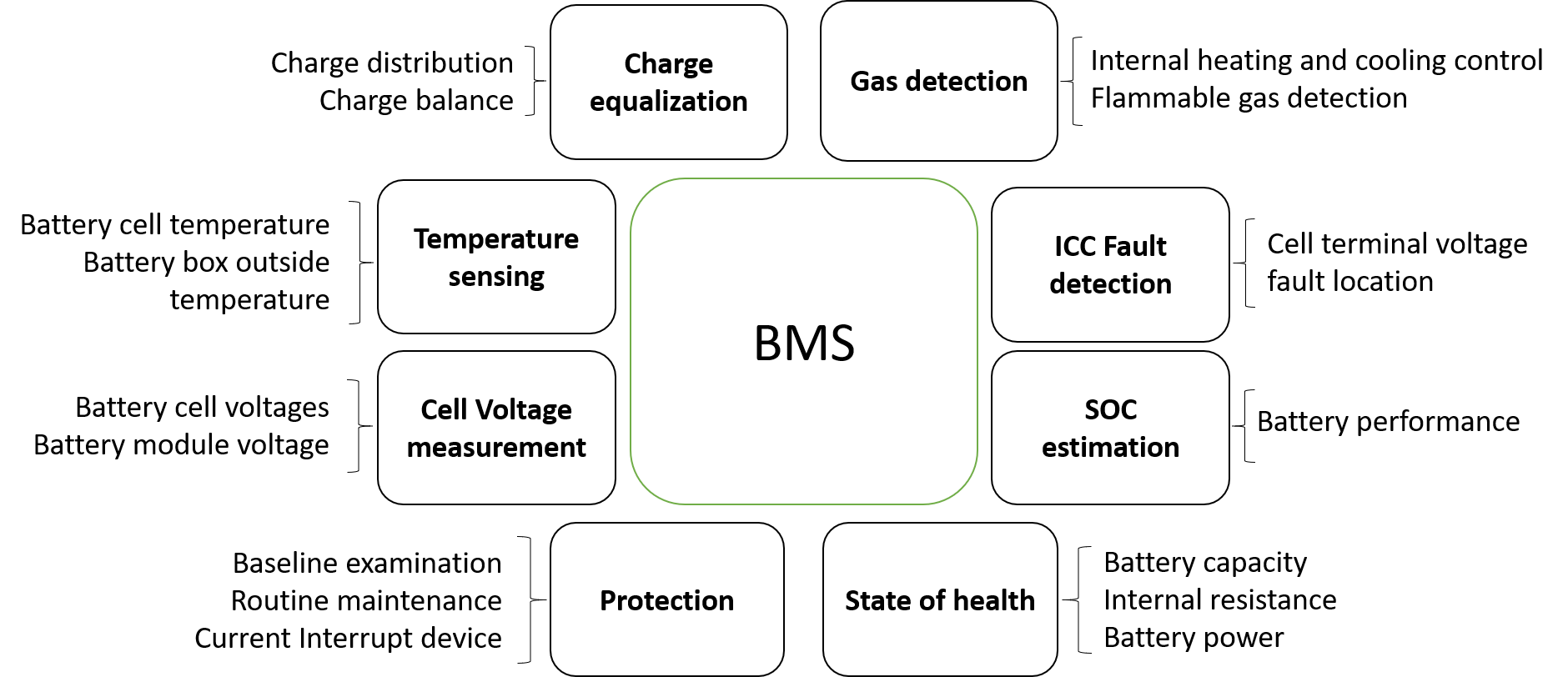}}
\caption{Overview of a Battery Management System for Hyperloop pod}
\label{fig}
\end{figure}
\subsection{Estimation of SOC for LiFePO4}
One of the functions of the BMS is to monitor the cell's performance, and its state of health for each battery cell. The use of the LiFePO4 batteries and the high voltage of the total battery pack necessitate the need for a more specialized BMS. Traditional SOC estimation algorithms cannot work accurately since these cells require more precise measurements of voltage and current due to the plateau in the Open Circuit Voltage - SOC curve. This curve is highly influenced by factors like discharge current rate, temperature, and aging state. Different methods for SOC predictions include the use of Neural Networks and fuzzy logic, but there are drawbacks of these as the accuracy of these algorithms is highly dependent on the model and inputs~\cite{languangkeyissues}. In this system, use of a Dual Extended Kalman Filter (DEKF) has been proposed to estimate the SOC~\cite{wangSOC}. This model is highly resistant to noise which is necessary when maintaining a high precision system. This system is also influenced by the charging voltage to eliminate any practical noise issues that might be encountered. With this, the accuracy of the system is higher than other algorithms. The high stability of the LiFePO4 batteries require more reliant SOC estimation, making the DEKF more preferred. The tests done using dynamic discharge and constant current show that the margin of error is less than 3\%.

\subsection{ICC Fault detection method} 

While dealing with High Voltage Battery packs like the one used in this system, it is essential to find any faults in individual cells instantaneously to avoid arcing. The Interclass Correlation Coefficient (ICC)~\cite{ICC} method is used to efficiently determine the strength of cells in the pack and enhances the safety and stability of the battery system. This method uses the real-time voltage data for instantaneous fault detection, consequently reducing the BMS computation load~\cite{ICCoriginal}. 

The ICC methodology is employed to examine battery faults by considering the unusual voltage dip. The coefficient value has excellent fault resolution as it magnifies the variation in voltage. Furthermore, the suggested method has a modest computational cost, thus lessening the redundant hardware design. The ICC equation is calculated using:

\begin{equation}
R_{i, 1} \\
={\left(x_{i, 1}-\frac{1}{2 n} \sum_{i=1}^{n}\left(x_{i, 1}+x_{i, 2}\right)\right)}\label{eq}\\
\end{equation}
\begin{equation}
R_{i, 2} \\
={\left(x_{i, 2}-\frac{1}{2 n} \sum_{i=1}^{n}\left(x_{i, 1}+x_{i,2}\right)\right)}\label{eq}\\
\end{equation}
\begin{equation}
 I C_{\left(x_{1}, x_{2}\right)} \\
=\frac{2\sum_{i=1}^{n}R_{i, 1}R_{i, 2} }{\sum_{i=1}^{n}R_{i, 1}^{2}+\sum_{i=1}^{n}R_{i, 2}^{2}}\label{eq}\\
\end{equation}
where, $x_{i,1}$ and $x_{i,2}$ represent the cell voltages in sampling time $i$.

Typically, the end-point voltages of two adjoining cells are similar under various driving cycles. An abrupt voltage dip forms the fault signals among two batteries at 1/100th sampling point. The sampling speed affects fault detection time to a certain extent. In case of a short circuit, there is a sharp fall in the terminal voltage within seconds. On detection of any fault in the batteries, a high priority emergency signal is sent to the CAN-BUS, which is received by the Vehicle Controller. The Vehicle Controller ,then, executes an Emergency Braking Protocol~\cite{ICC}. 

This cited method can be further improved upon by using an adaptive sensing model. As shown in Fig.5, an adaptive threshold and sampling rate can be set based on different values of motor RPM, velocity, and acceleration of the Hyperloop pod, since large values of these parameters increase the likeliness of fault occurence. These parameters can be directly obtained from the outputs of the pose estimation algorithm, and vehicle controller node.

\begin{figure}[htbp]
\centerline{\includegraphics[width = 244pt]{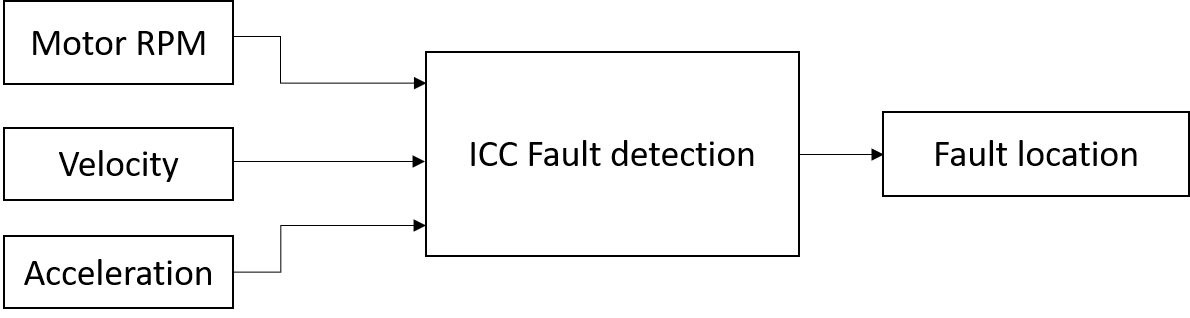}}
\caption{ICC fault detection flow diagram}
\label{ICCbased}
\end{figure}
\subsection{Protection circuit}
Preventing an arc flash can be done by taking precautionary measures such as baseline examination and assuring routine maintenance of the electric device. Another preventive measure is to de-energize the devices. However, there is a need for other protection circuits in case other preventive measures fail. There can be several reasons for battery breakdowns, such as over-discharge, overcharge, overheating, and internal short circuits. The main characteristics of runaway are excess gas pressure and overheating. One way to restrict these consequences is having a suitable cell design, with safety provisions like venting and insulation layers. Most 18650 cells contain protections like a current interrupt device (CID), having a gas vent, and a current interrupter. CID is used to control cell overcharge by unfolding the cell circuitry by internal pressure rise. Some commercial Li-ion batteries incorporate a Positive Temperature Coefficient (PTC), which is used to prevent high currents inside the cell. The high voltage battery packs on the Hyperloop Pod require numerous cells, and controlling the current into the cells is essential.  

The IC BQ76940, manufactured by Texas Instruments, fulfills this function of a CID. The main reason for arcing is a high discharge of the cells. The IC serves as a distribution board where multiple cell terminals can be connected to it. It helps to distribute charge, in case any cell is discharging at unusual rates, thereby preventing the possibility of leakage and arcing.

\section{Pose Estimation}

In this section, a pose estimation model is proposed which utilizes only on-board computing sensors and already present tube structures. To fuse estimates, a Kalman filter has been used for simplicity which can be extended into an Extended Kalman Filter (EKF) or Unscented Kalman Filter (UKF) in case of non-linear dynamics. A brief overview of the Kalman filter is given, followed by the algorithms required for each of the four parameters, namely, acceleration, velocity and position.

\subsection{Introduction to the Kalman Filter}
The Kalman filter in essence, is a naive Bayesian filter with some assumptions about the dynamics and noise.  It assumes the dynamic model for the pose estimation is linear and noise is Gaussian. The Kalman Filter works broadly in two steps, a prediction step and an update step. During the prediction step, a belief of the parameter estimate is calculated through equations of motion of the dynamic model. The parameter belief and output estimates are expressed in Gaussian terms, that is a mean and a co-variance matrix. Following this, an update step considers measurements from another source and outputs the parameter estimates. A good estimate should have low bias from the ground truth as well as have low co-variance, in other terms, the centroid should be close to the ground truth while the spread of the Gaussian distribution should be narrow/densely packed. The following flowchart illustrates the KF working.

\subsection{Usage in the Hyperloop Pod}
For the hyperloop pod, the algorithm pseudo-codes for each parameter have been shown below. The models considered are simple models that use the kinematic equations of the pod. For simplicity, motion is assumed to be in a straight line along the x-direction only. All sensor noises are assumed to have Gaussian distribution and constant.
Since, the IMU BNO-055 already outputs filtered orientation values, a simple averaging filter is used for giving the orientation estimate. 
For the acceleration model, the prediction step is carried out by taking the value from the acceleration-time graph at discrete time intervals. For the update step, the acceleration is measured from the IMU. The output after the update step returns a mean updated value of the acceleration along with the updated co-variance.

\begin{equation}
 A_{u} \rightarrow A-t \text { curve time update} \label{eq}\\   
\end{equation}
\begin{equation}
Z_{t}=I A_{t}\label{eq} \\
\end{equation}
\begin{equation}
 A_{c o v}=R_{t}\label{eq} \\   
\end{equation}
\begin{equation}
K_{t}=A_{c o v} I\left(I A_{c o v} I+Q_{t}\right)^{-1}\label{eq}\\
\end{equation}
\begin{equation}
 A_{n e w}=A_{u}+K_{t}\left(Z_{t}-I A_{u}\right)\label{eq} \\   
\end{equation}
\begin{equation}
  A_{n e w_{c o v}}=\left(I-K_{t} I\right) A_{c o v}\label{eq} \\  
\end{equation}

The velocity model does not require simulated data, however, considers the acceleration estimate obtained in the previous step. The prediction step is carried out through simple equations of motion and involves the acceleration estimate output. For the update step, the sensor model includes the tachometer output multiplied with the transformation matrix that provides measurement of the velocity. The velocity estimate output is computed in the form a mean and covariance matrix for a normal distribution.

\begin{equation}
 V_{u(t)}=I V_{u(t-1)}+I(\Delta t) A_{t}\label{eq}\\
 \end{equation}
 \begin{equation}
Z_{t}=\frac{1}{R} * V_{t}\label{eq}
\end{equation}
\begin{equation}
V_{c o v(t)}=I V_{c o v(t-1)} I+R_{t}\label{eq}
\end{equation}
\begin{equation}
K_{t}=V_{c o v(t)} I\left(I V_{c o v(t)} I+Q_{t}\right)^{-1}\label{eq} \\
\end{equation}
\begin{equation}
V_{n e w}=V_{u(t)}+K_{t}\left(Z_{t}-\frac{1}{R} \cdot V_{u(t)}\right)\label{eq} \\
\end{equation}
\begin{equation}
 \qquad V_{n e w_{c o v}}=\left(I-\frac{1}{R} \cdot K_{t}\right) V_{c o v(t)}\label{eq}
\end{equation}
 
Finally, the position model is calculated that considers displacement from the start of the track.  For the prediction step, the position is calculated using a simple kinematic equation of motion taking the velocity estimate and acceleration estimate. The update step involves the pre-multiplication with the transformation matrix with the optical encoder output. The output received is the position from start along the direction of motion and is expressed as a normal distribution.

\begin{equation}
X_{u(t)}=I X_{u(t-1)}+\Delta t \quad 0.5(\Delta t)_{A_{t}}^{V_{t}}\label{eq}\\  
\end{equation}
\begin{equation}
X_{u(t)}=I X_{u(t-1)}+\Delta t \quad 0.5(\Delta t)_{A_{t}}^{V_{t}}\label{eq}\\    
\end{equation}
\begin{equation}
Z_{t}=\frac{N}{2 \pi R} * X_{t}\label{eq}\\   
\end{equation}
\begin{equation}
X_{\operatorname{cov}(t)}=I X_{\operatorname{cov}(t-1)} I+R_{t}\label{eq}\\   
\end{equation}
\begin{equation}
K_{t}=X_{\operatorname{cov}(t)} I\left(I X_{\operatorname{cov}(t)} I+Q_{t}\right)^{-1}\label{eq}\\   
\end{equation}

\begin{equation}
X_{\text {new}}=X_{u(t)}+K_{t}\left(Z_{t}-\frac{N}{2 \pi R} * X_{u(t)}\right)\label{eq}\\   
\end{equation}

\begin{equation}
X_{\text {new}_{\text {cov}}}=\left(I-\frac{N}{2 \pi R} * K_{t}\right) X_{\operatorname{cov}(t)}\label{eq}\\  
\end{equation}

It is to be noted that fiducial markers serve as updates for position as well as re-calibration markers for Kalman gains.

\section{Results}

To ensure that the approaches proposed in this paper are suitable for application in practical systems, this work further includes results from software simulations performed in Simulink and Python.

\begin{figure}[htbp]
\centerline{\includegraphics[width = 244pt]{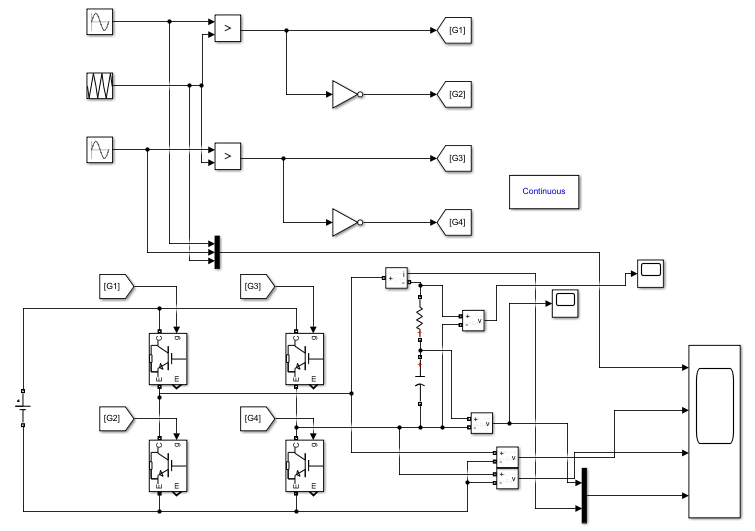}}
\caption{Simulink Model of DC-to-AC inverter circuit}
\label{fig:simulink}
\end{figure}

\begin{figure}[htbp]
\centerline{\includegraphics[width = 244pt]{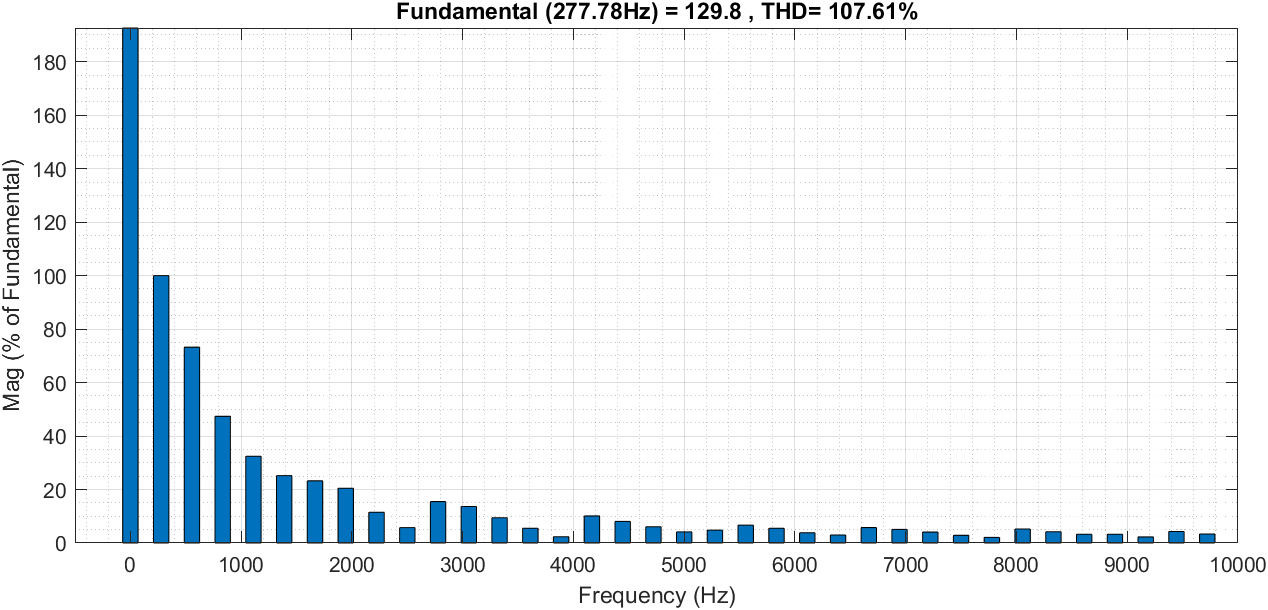}}
\caption{Time Domain FFT plot of output waveform from the designed inverter}
\label{fig:timedomainfft}
\end{figure}

\begin{figure}[htbp]
\centerline{\includegraphics[width = 244pt]{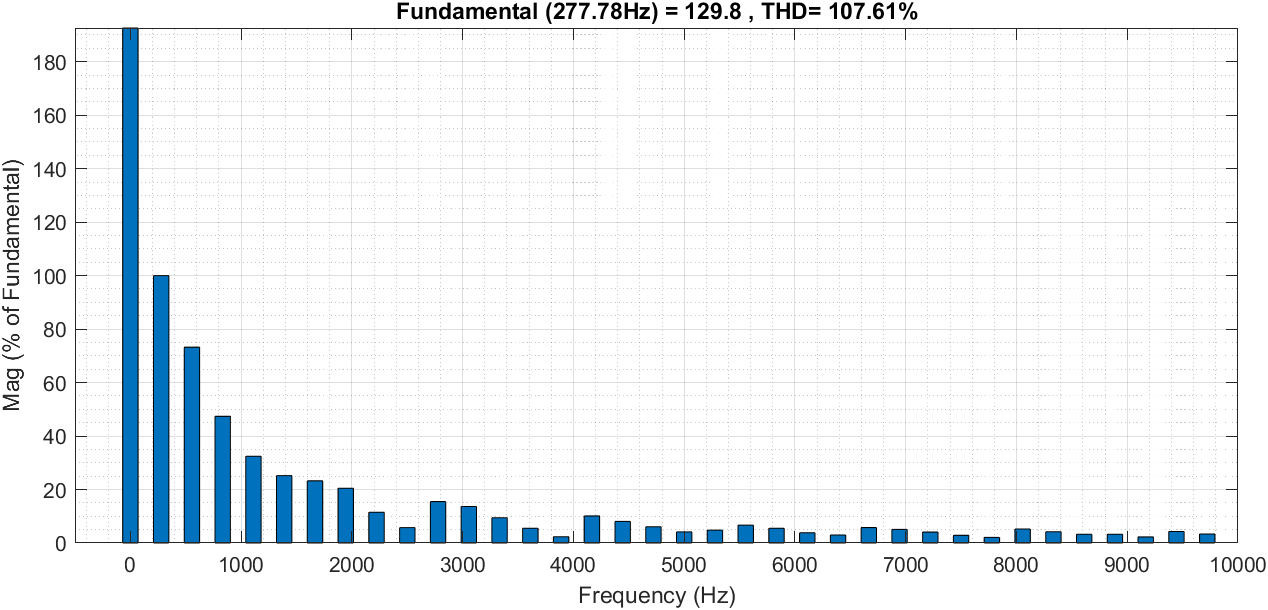}}
\caption{Frequency Domain FFT plot of output waveform from the designed inverter}
\label{fig:freqdomainfft}
\end{figure}

To evaluate the performance of the DC-to-AC inverter model (refer Fig~\ref{fig:simulink}), the circuit was designed from scratch using synchronous insulated-gate bipolar transistors (IGBT) as switches. The input DC voltage of 350V was taken directly across the terminals of the HV battery pack. Further removal of unwanted harmonics in output waveform was achieved by appropriate low pass filtering using a $1k\Omega$ resistor, and a $40nF$ capacitor. Ideal AC frequency for operation of practical LIMs has seen to around $277.77Hz$, as mentioned in works like~\cite{epfl}. The inverter design outputs this exact frequency with highest magnitude (refer Fig~\ref{fig:timedomainfft} and Fig~\ref{fig:freqdomainfft}). The output DC voltage can further be clipped off or completely eliminated (using a high pass filter) based on the DC voltage necessary for the specific LIM propulsion design implemented.

With respect to the electronics system design, to make the pod design scalable, a Communication Unit is to be added to the System Architecture to allow real-time monitoring of pod health and location from the base station. This will also allow the base station to stop the pod remotely, if necessary. The existing model can also be expanded by adding redundancy in sensors and Units, and adding provision to switch to a backup board, in case of any fault in the primary board. The next step, after the design stage, will be to test the design on physical hardware and finally to evaluate its performance when integrated with the electrical and mechanical subsystems.

For pose estimation, a python-based simulation was carried out for each of the acceleration, velocity, and displacement parameters. Each parameter was simulated for approximately 120 discrete time-steps. The graph shapes were taken from the University of Colorado Boulder 2019 report. The loss function that has been used is the root mean-squared error (RMSE).

\begin{equation}
\label{eq:rmse}
\begin{split}
RMSE = \sqrt{\frac{\sum(x_{pred}-x_{real})}{n}}
\end{split}
\end{equation}

Where $xpred$ and $xreal$ are the predicted (after Kalman filtering) and real values. $n$ is the total number of samples, which is 120 in this case.

\begin{figure}[htbp]
\centerline{\includegraphics[width=244pt]{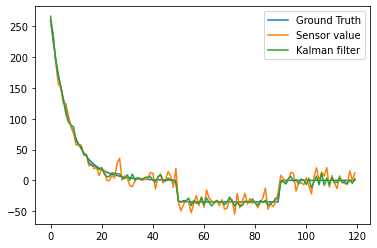}}
\caption{Acceleration-time filtering}
\label{fig:cA}
\end{figure}

For acceleration, the ideal or ground-truth curve was approximated by scaling down a base-10 logarithmic curve followed by scaled step-functions for the negative acceleration period. As is visible in Fig~\ref{fig:cA}, the peak occurs at the beginning at 260m/s2. The measurement noise was assumed to be Gaussian in nature and was sampled 120 times with a scaled standard deviation of 10. Similarly, the process noise was defined, however, with a standard deviation of 5. A one-dimensional Kalman filter was then used. It was seen that the RMSE for raw sensor output was 53.84 whereas the RMSE for Kalman filtering was just 4.44.

\begin{figure}[htbp]
\centerline{\includegraphics[width=244pt]{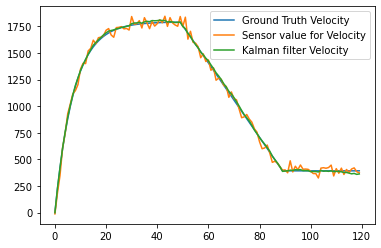}}
\caption{Velocity-time filtering}
\label{fig:V}
\end{figure}

This predicted value was then used for the velocity prediction step. The process and measurement noise were again defined as gaussian with standard deviations of 35 and 10.Drawing an inference from Fig~\ref{fig:V},it was seen that the RMSE of the raw sensor output was 35.97 compared to an RMSE of 11.73 with the Kalman filter.

\begin{figure}[htbp]
\centerline{\includegraphics[width=244pt]{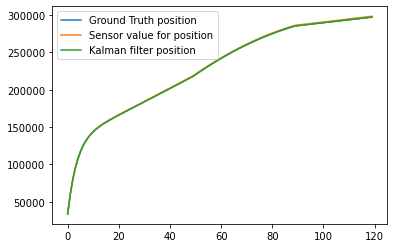}}
\caption{Position-time filtering}
\label{fig:P}
\end{figure}
For the position parameter, a similar Kalman filter was used, however the RMSE was higher and comparatively inconsistent as the acceleration and velocity estimates as is visible from Fig~\ref{fig:P}. This however, can be attributed to the fact that the scales for each parameter vary in terms of the power of 10. Hence, relatively speaking, the Kalman filter was successful in keeping the percentage RMSE very low. Moreover, in the real world, re-calibration of the sensors and Kalman gains happens with crossing of every fiducial marker as well as end points. Hence, this would essentially reduce the error further.

\section*{Conclusion}

The board layouts presented in this work, along with sensor interfacing and intra-board communication, showcase a simple, yet thorough and distributed system design that covers all the sensing and monitoring needs of the pod. Extensive literature review and analysis of battery management, protection, and fault detection schemes led to the final proposal of an efficient and self-sufficient battery monitoring system. Further simulation results verified the feasibility and performance of accurate real time position estimation algorithms, and electric system components.

\vspace{12pt}
\color{red}

\end{document}